# Point-defects assisted Zn-diffusion in AlGaInP/GaInP systems during the MOVPE growth of inverted multijunction solar cells

Manuel Hinojosa, Iván García*, Shabnam Dadgostar[2] and Carlos Algora[1]

[1]Instituto de Energía Solar – Universidad Politécnica de Madrid; [2] OptronLab – Universidad de Valladolid

*igarcia@ies.upm.es

*Abstract*— We investigate the dynamics of Zn diffusion in MOVPE-grown AlGaInP/GaInP systems by the comparison of different structures that emulate the back-surface field (BSF) and base layers of a GaInP subcell integrated into an inverted multijunction solar cell structure. Through the analysis of secondary ion mass spectroscopy (SIMS), electrochemical capacitance-voltage (ECV) and spectrally resolved cathodoluminescence (CL) measurements, we provide experimental evidence that 1) the Zn diffusion is enhanced by point defects injected during the growth of the tunnel junction cathode layer; 2) the intensity of the process is determined by the cathode doping level and it happens for different cathode materials; 3) the mobile Zn is positively charged and 4) the diffusion mechanism reduces the CuPt ordering in GaInP. We demonstrate that using barrier layers the diffusion of point defects can be mitigated, so that they do not reach Zn-doped layers, preventing its diffusion. Finally, the impact of Zn diffusion on solar cells with different Zn-profiles is evaluated by comparing the electrical I-V curves at different concentrations. The results rule out the introduction of internal barriers in the BSF but illustrate how Zn diffusion under typical growth condition can reach the emitter and dramatically affect the series resistance, among other effects.

## I. INTRODUCTION

Zinc (Zn) is the most common p-type dopant element in III-V materials grown by metal-organic-vapor-phase-epitaxy (MOVPE). However, Zn has proved to be a strong and rapid diffuser, which complicates the achievement of intended doping profiles and can spoil the optoelectronic performance of a variety of devices such as bipolar junction transistors (BJT), light-emitting diodes (LED), lasers or multijunction solar cells (MJSC) [1][2][3][4][5]. Previous investigations revealed that the n-type doping of nearby layers enhances the Zn diffusion process [6][7]. Particularly, the widely accepted model proposed for an AlGaAs/GaAs system by Deppe et al. suggests that the Fermi level pinning at the surface during the growth of a heavily doped n-type GaAs leads to an imbalance and injection of group-III interstitials ($Ga_I$) [8]. According to the model, the injected defects diffuse rapidly from the surface to buried layers, where active Zn is located in the group-III site of the lattice ($Zn_{Ga}^-$), and promotes the so-called kick-out process. This way, $Ga_I$ transfers Zn to an interstitial position ($Zn_I$) and reduces the local concentration of holes [$h^+$]. Finally, the interstitial Zn diffuses, and the reverse process takes place via Frenkel mechanisms, releasing new Ga interstitials.

$$Ga_I + Zn_{Ga}^- + h^+ \leftrightarrow Ga_{Ga} + Zn_I \qquad (1)$$

Thus, the growth of a highly n-doped layer on the top of a Zn-doped AlGaInP/GaInP system is expected to induce a redistribution of Zn, group-III species and holes across the structure by means of a mechanism that can be divided into five stages: 1) the injection of group-III interstitials during the growth of the n-type layer, 2) the diffusion of these defects from the surface to the buried Zn-doped layers, 3) the kick-out mechanism, which transfers Zn from a substitutional to an interstitial position, 4) the diffusion of interstitial Zn and, finally, 5) the Frenkel reaction, which pushes the Zn back to a group-III lattice site and releases new interstitial group-III atoms. On the other hand, the bandgap in a GaInP alloy highly depends on the spontaneous CuPt ordering produced by the arrangement in alternate (111) planes of the group-III sublattice atoms during the growth [9]. The Zn diffusion in an AlGaInP/GaInP system implies the transfer of group-III elements to interstitial positions −equation (1), from right to left−, with subsequent kick-out processes −equation (1), from left to right. As a consequence, the constant movement of group-III species throughout different positions of the crystal lattice favors the disruption of the metastable arrangement of Ga and In atoms present in ordered GaInP and thereby can modify the bandgap of the GaInP layer [10].

The incorporation of highly doped n-type layers in III-V structures is required in most optoelectronic applications. For example, in a MJSC the different bandgap subcells are monolithically integrated through optically transparent tunnel diodes, also termed tunnel junctions (TJ). A TJ is essentially a thin *pn* junction that exhibits a low resistive behavior as long as the electrical conduction is assisted by the tunnel effect. These thin layers enable the interconnection of different subcells with negligible electrical losses. In order to obtain a high peak current and a low resistivity, highly doped cathode and anode layers, with typical doping densities exceeding $1\cdot 10^{19}$ cm$^{-3}$, are required [11]. Hence, the integration of tunnel junctions in the semiconductor structure represents an unavoidable injection of point defects that can potentially interact with the electrically active Zn found in already grown surrounding layers. For instance, in inverted multijunction solar cells, the TJ's are grown immediately after the passivating back surface field (BSF) layers of the subcells, where concentrations of Zn around $1\cdot 10^{18}$ cm$^{-3}$ are typically found, turning out the diffusion of Zn almost unavoidable [12][13]. In this work, we investigate the dynamics of Zn diffusion in different p-AlGaInP/p-GaInP systems corresponding to the TJ-cathode/barrier/BSF/absorber layers of a GaInP subcell integrated into a MJSC system. The baseline structure used in the study, represented in Fig. 1, mimics the layer arrangement in an inverted GaInP solar cell with a tunnel junction grown on top. In previous works we observed that the AlGaAs:C anode of the TJ does not play a significant role in the Zn diffusion mechanism. For simplicity and to prevent complications during the etching steps of the capacitance-voltage measurements (ECV) of the cathode layer, the anode is not included in the baseline structure. From this generic structure, we modify systematically structural and growth parameters to evaluate their impact on the [Zn] and [h$^+$] profiles. In particular we focus on: 1) the influence of the doping level used in the cathode of the TJ; 2) the use of different TJ cathode materials, 3) the use of (Al)(Ga)InP alloys with different Al content in the BSF and, finally, (4) the effectiveness of using different barrier layers to stop the propagation of point defects from the cathode to the BSF. The effect of Zn diffusion on the GaInP bandgap are further analyzed by using the cathodoluminescence (CL) technique.

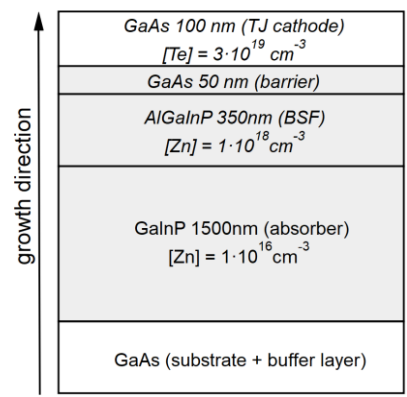

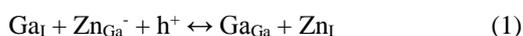

Fig. 1 Sketch of the baseline structure with the baseline doping levels used. The parameters varied in this structure are: (1) the doping density in the cathode of the TJ, (2) the cathode material used in the TJ, (3) the Al content in the BSF layer and, finally, (4) the material and thickness of the barrier layer. The GaInP absorber is identical in all cases.

## II. Experimental

All samples were grown on GaAs substrates with a 2º miscut towards the (111)B plane in a horizontal low-pressure MOVPE reactor (AIX200/4) at 100 mbar. The precursors used were AsH$_3$ and PH$_3$ for group-V, TMGa and TMIn for group-III and DETe and DMZn for dopant elements. The GaAs cathode layer was grown at 550 ºC, with a deposition rate of 2.1 µm/h and the V/III ratio was 20. The GaInP, AlGaInP and AlInP layers were grown at 675 ºC, 4 µm/h and with V/III ratios of 110, 110 and 140, respectively. The partial pressures of DMZn and DETe used to obtain p-type and n-type electrical doping are specified in Table I. The DMZn used in the absorber and the BSF layers was kept constant in all samples, intending to attain doping levels of ~$1\cdot 10^{16}$ cm$^{-3}$ and ~$1\cdot 10^{18}$ cm$^{-3}$, respectively. On the other hand, the barrier layers were nominally undoped. An ex-situ MOVPE thermal annealing process of 60 minutes at 675 ºC under AsH$_3$ environment was conducted in all samples in order to reproduce the thermal load of a complete epitaxial process of a 2-junction solar cell. The results presented correspond to the annealed samples unless the contrary is specified. Solar cell devices were fabricated using the inverted metamorphic (IMM) solar cell fabrication process as described elsewhere [14] with an active area of 0.1 cm$^2$. Both front and rear contacts are based on ~300 nm of gold deposited by electroplating. All the layers are lattice-matched to GaAs, as confirmed by the high-resolution X-ray diffraction (HRXRD) measurements on the Ga$_{0.5}$In$_{0.5}$P, Al$_{0.2}$Ga$_{0.3}$In$_{0.5}$P and Al$_{0.5}$In$_{0.5}$P layers. The free-carrier and Zn concentrations depth profiles were measured by ECV and secondary ion mass spectroscopy (SIMS) techniques. In the SIMS profiles, all layers were delineated through the detection of Al, In, P and As. The donor concentration in the cathode layer (N$_D$) was obtained through the electron concentration (n) measured by ECV. The growth conditions favor the formation of a high CuPt ordering in the group-III sublattice of GaInP, which can be disrupted by the diffusion

TABLE I. Growth details of the samples

| ID | Cathode | P$_{DETe}$ cathode (mbar) | [N$_D$] cathode (cm$^{-3}$) | Barrier | BSF |
|---|---|---|---|---|---|
| A | GaAs | 0 | - | 50 nm GaAs | AlGaInP |
| B | GaAs | $2.05\cdot 10^{-6}$ | $8\cdot 10^{18}$ | 50 nm GaAs | AlGaInP |
| C | GaAs | $1.03\cdot 10^{-5}$ | $1\cdot 10^{19}$ | 50 nm GaAs | AlGaInP |
| D | GaAs | $2.05\cdot 10^{-5}$ | $3\cdot 10^{19}$ | 50 nm GaAs | AlGaInP |
| E | AlInP | 0 | - | 50 nm GaAs | AlGaInP |
| F | AlInP | $2.57\cdot 10^{-5}$ | $5\cdot 10^{18}$ | 50 nm GaAs | AlGaInP |
| G | GaAs | $2.05\cdot 10^{-5}$ | $3\cdot 10^{19}$ | 500 nm GaAs | AlGaInP |
| H | GaAs | $2.05\cdot 10^{-5}$ | $3\cdot 10^{19}$ | 50 nm AlAs | AlGaInP |
| I | GaAs | $2.05\cdot 10^{-5}$ | $3\cdot 10^{19}$ | 50 nm AlInP | AlGaInP |
| J | GaAs | $2.05\cdot 10^{-5}$ | $3\cdot 10^{19}$ | 50 nm GaAs | GaInP |
| K | GaAs | $2.05\cdot 10^{-5}$ | $3\cdot 10^{19}$ | 50 nm GaAs | AlInP |

*The thickness of the cathode, BSF and absorber layers are 100 nm, 350 nm and 1500 nm.

**The DMZn partial pressure used during the growth of the BSF and absorber layers was identical in all samples: P$_{DMZn}$ = $1.02\cdot 10^{-2}$ mbar in the BSF and P$_{DMZn}$ = $2.8\cdot 10^{-5}$ mbar in the absorber.

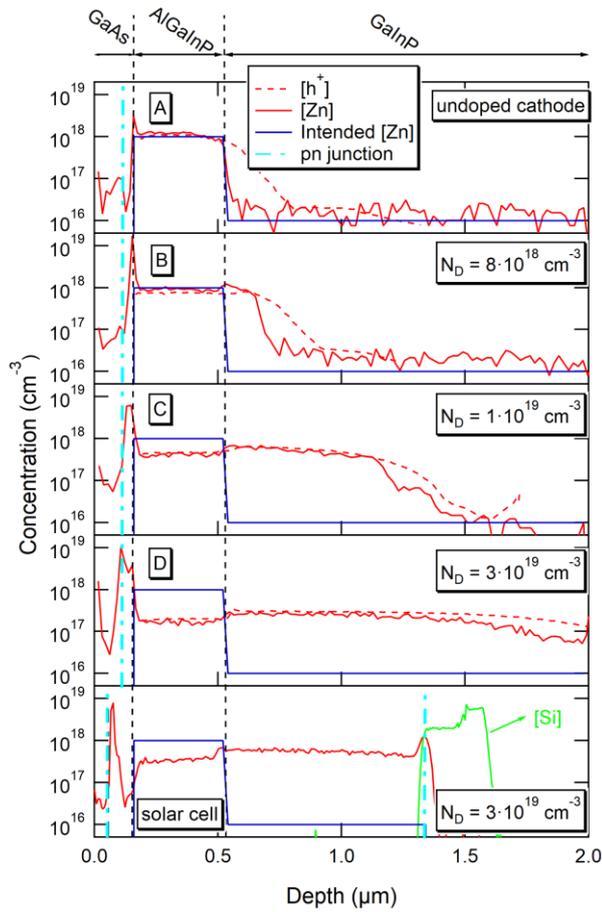

Fig. 2. Depth profiles of [Zn] and [h+] in the AlGaInP/GaInP layers of samples A, B, C, D −with increasing donor concentration in the GaAs cathode layer− and a reference solar cell incorporating a TJ with the same cathode as D. The position of pn junctions in each structure is also indicated.

of Zn. This was analyzed by spectrally resolved plan view luminescence scans from the AlGaInP/GaInP (BSF/absorber) structures, collected at 83 K by the cathodoluminescence (CL) technique using a monoCL 2 (Gatan UK) CL system attached to a LEO 1530 (Carl-Zeiss) field-emission scanning electron microscope (FESEM), using e-beams of 5 kV. The details of the samples are summarized in Table I. Finally, I-V curves under concentration were measured in the solar cells using a custom made, flash-lamp based, setup.

## III. RESULTS

### A. Effect of the cathode doping level on Zn diffusion

First, we focus on the influence of the n-type doping level in the GaAs cathode on the diffusion of Zn along an AlGaInP/GaInP system. For this, we use samples A, B, C and D, with identical structures and growth conditions, but different [Te] in the cathode. [Zn] and [h+] depth profiles are shown in Fig. 2. In general, the ECV profiles match the SIMS profiles, but some differences are found in the transition regions. This is due to the spatial resolution of the ECV measurement, which integrates the depletion region width at each point. Thus, the spatial resolution fluctuates between 20 and 200 nm depending on the concentration level. The doping attained in the AlGaInP and GaInP layers matches the desired profile in the absence of a doped cathode (A). In contrast, the [Zn] and [h+] profiles become dramatically altered as the GaAs layer is increasingly doped (B, C and D). As a result of the diffusion process, the [h+] profile flattens. The position of the diffusion front inside the GaInP absorber, together with the drop of the [Zn] level from the nominal in the AlGaInP BSF, reflects the intensity of the process and shows the correlation with the donor concentration in the cathode layer. For instance, in the most severe diffusion case (D, with $N_D = 3·10^{19}$ cm$^{-3}$), [Zn] has already spread all around the whole absorber GaInP layer and the level in the BSF layer has gone down from $1·10^{18}$ cm$^{-3}$ to $2·10^{17}$ cm$^{-3}$. The altered profiles reveal that the growth of a highly doped n-type GaAs layer after the BSF, together with the presence of a concentration gradient and certain thermal energy, promotes the diffusion of Zn. Otherwise, the diffusion of Zn is insignificant. This dynamic differs from a conventional diffusion process and suggests 1) the assistance of point defects injected during the growth of the TJ cathode, at an intensity related to the doping level, which activates the diffusion mechanism, and 2) different diffusivities of Zn depending on its position in the crystal lattice (substitutional or interstitial). According to the kick-out reaction, the injected defects that reach the BSF layer transfer the electrically active Zn from a substitutional acceptor position −group-III sublattice− to a mobile interstitial position (equation (1), from left to right). As long as the density of the injected interstitial Ga depends on the electron concentration in the cathode, it becomes a key parameter to determine the intensity of the process: a higher $N_D$ in the cathode certainly induces a stronger and faster diffusion of Zn.

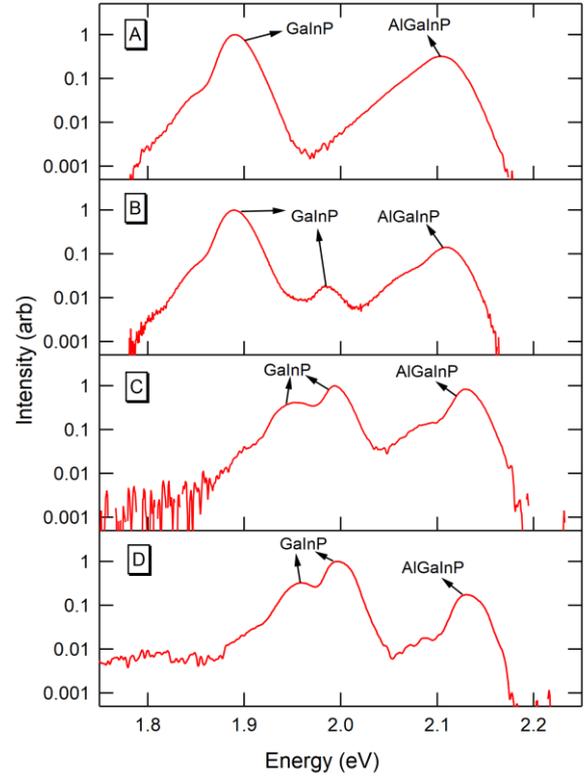

Fig.3. CL spectral emission of samples A, B, C and D. The peaks in CL are labeled identifying the layer where the emission is originated: the absorber GaInP and the BSF AlGaInP.

The accumulation of Zn that appears in the GaAs layer, close to the GaAs/p-AlGaInP interface coincides with the np junction formed by the n-type cathode layer. This indicates that the propagation of Zn is stopped by the built-in electric field of this *pn* junction. In order to provide further insight on this point, the [Zn] profile of an inverted GaInP solar cell integrated into a MJSC is included in Fig. 2. The dopant flows used in the cathode of TJ and the BSF layers correspond to case D, but the GaInP absorber layer comprises a 750 nm Zn-doped and a 180 nm Si-doped layers that form a deep *pn* junction in the absorber layer. It can be seen that the Zn-redistribution is similar as in case A (demonstrating that the TJ anode plays no significant role), but the Zn coming from the BSF is stopped at the Si-doped emitter region. Note that the junction at the TJ is a bit shifted to the left in this case because of the presence of the anode layer. The donor concentration in the n-side close to the GaAs/AlGaInP interface ($N_D = 3 \cdot 10^{19}$ cm$^{-3}$) is much higher than in the absorber junction ($N_D = 1 \cdot 10^{18}$ cm$^{-3}$), so the magnitude of the electric field and the ability to stop charged species differs significantly. The difference in the level of [Zn] detected at both interfaces suggests an interaction between the internal electric fields and the mobile Zn: apparently, the higher the electric field, the higher the Zn peak concentration. In addition, the direction of the electric fields, from the n-layer to the p-layer, reveals that the mobile Zn is positively charged and, hence, occupies interstitial positions during the propagation. Similar accumulation of Zn at *pn* junctions have been observed elsewhere [5][6][15].

The plan view CL spectral emissions at 83 K shown in Fig.3, are used to evaluate the impact of Zn diffusion on the bandgap of the GaInP absorber layer. HRXRD scans confirmed that the composition of GaInP layers was virtually identical, so the bandgap differences can be directly related to modifications in the ordering state. In the absence of Zn diffusion (case A), the peak emission corresponding to the GaInP layer is produced at 1.89 eV. On the other hand, as the cathode is increasingly doped, the bandgap varies significantly. In B ($N_D = 8 \cdot 10^{18}$ cm$^{-3}$), the GaInP CL emission is composed of two peaks: a primary peak, located at a similar energy as in A (1.89 eV), and a second peak, located at a higher energy (1.99 eV). This value corresponds to the bandgap of the almost completely disordered GaInP at the CL measurement temperature [9]. In C and D, with a more drastic Zn diffusion ($N_D > 1 \cdot 10^{19}$ cm$^{-3}$), the main peak is located at 2 eV, with some low intense signal still detected at lower energies and a clear secondary peak at 1.95 eV. The absolute increase of the peak energy (110 meV) reveals a drastic reduction of the average CuPt ordering degree related to the intensity of the diffusion process. The bandgap variations induced by Zn diffusion affect the external quantum efficiency in the resulting GaInP solar cells, as shown in previous studies [12]. The emission of multiple peaks suggests the presence of regions with different ordering state in the GaInP absorber layer. In addition, the AlGaInP layers in samples C and D emit at ~20 meV higher energies than in A, which suggest differences on the ordering state in those layers too.

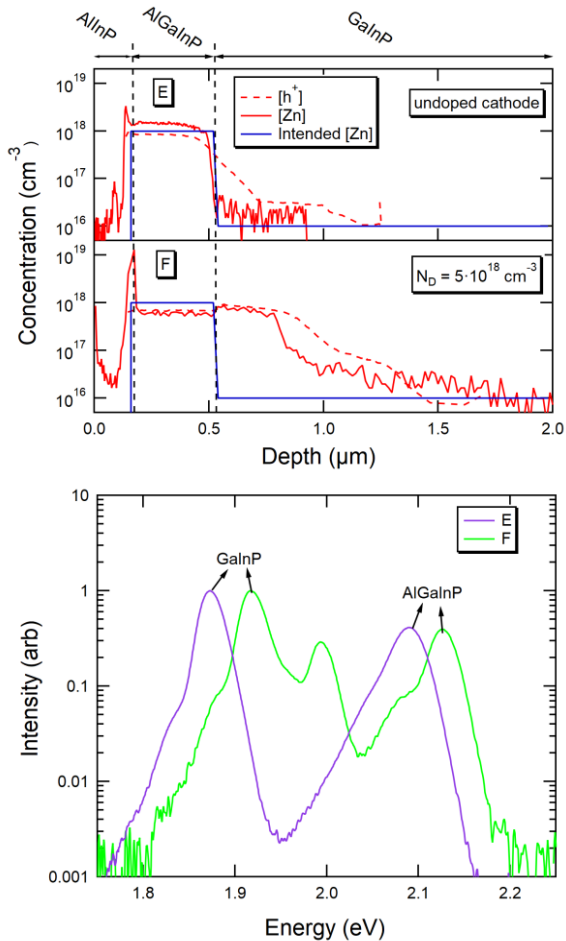

Fig. 4. Top: Depth profiles of [Zn] and [h+] in the AlGaInP/GaInP layers of Samples E and F —with doped and undoped AlInP cathode layers, respectively. Bottom: CL spectral emissions.

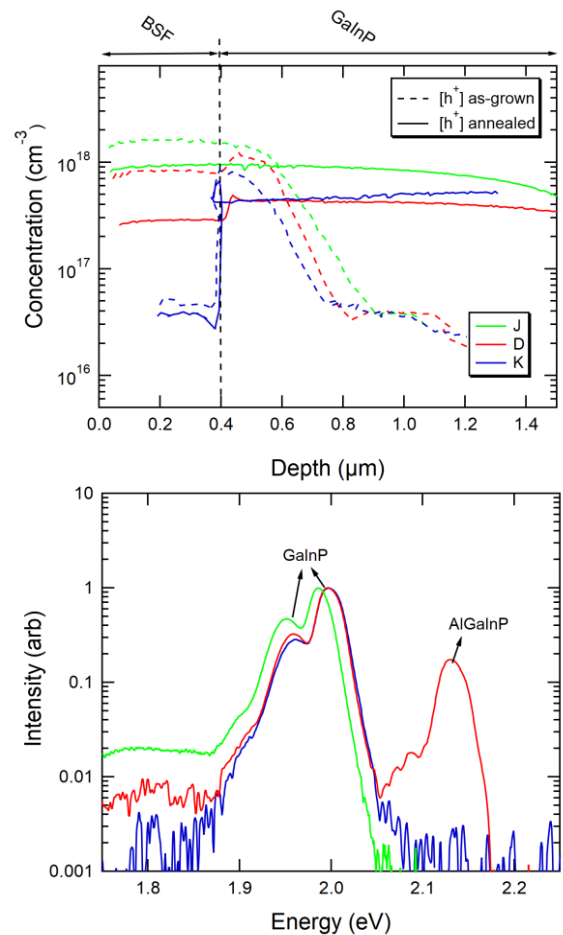

Fig. 5. Top: Depth profiles of [h+] in the AlGaInP/GaInP layers of as-grown and annealed Samples J, D and K —with BSF layers of GaInP, AlGaInP and AlInP, respectively. Bottom: CL spectral emissions of the annealed samples.

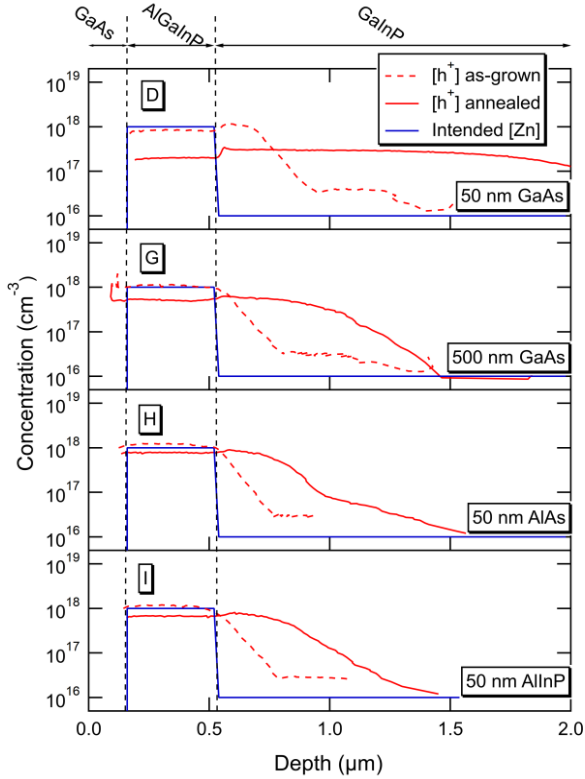

Fig. 6. Depth profiles of [h+] in the AlGaInP/GaInP layers of the as-grown and annealed Samples D, G, H and I, with varying barrier layers. The barriers are 50 nm GaAs (D), 500 nm GaAs (G), 50 nm AlAs (H) and 50 nm AlInP (I), as indicated with the labels.

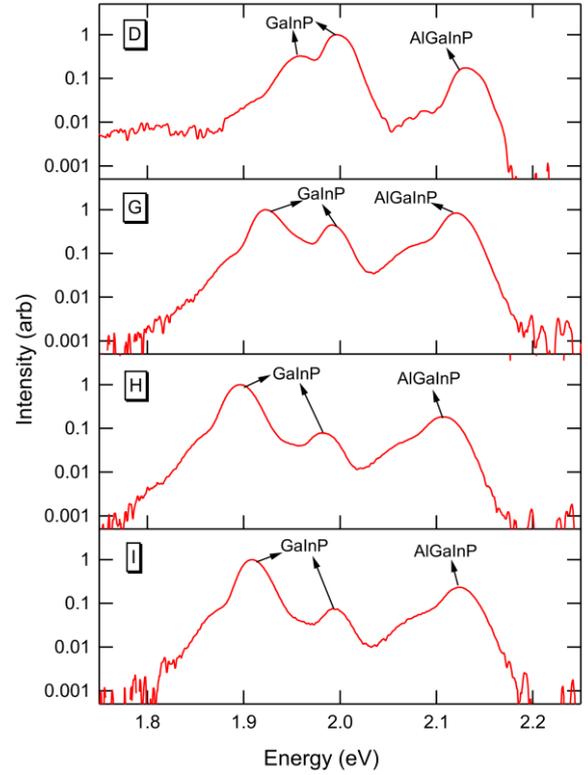

Fig. 7. CL spectral emission of annealed Samples D, G, H and I, with varying barrier layers. The barriers are 50 nm GaAs (D), 500 nm GaAs (G), 50 nm AlAs (H) and 50 nm AlInP (I).

*B. Effect of the cathode material on the injection of point defects and Zn diffusion*

Next, we focus on the influence of the material used as cathode in the TJ. Using alternative materials in this cathode might produce differences in the diffusion dynamics by modifying either the density or the nature of the injected point defects. Two structures using either an undoped (sample E) or a doped AlInP cathode (sample F) are evaluated, to test if group-III Al and In species can induce diffusion of Zn similarly as Ga. The concentration profiles and the CL spectra are shown in Fig. 4. Again, the same trends are reproduced: the doping in the cathode activates the diffusion of Zn along the AlGaInP and GaInP layers as in Fig. 3. Therefore, the mechanism described by Deppe applies for different III-V systems: the kick-out and Frenkel reactions in (1) can be extended to other III-V materials by changing the Ga species by the corresponding group-III element. In this case, interstitial Al and/or interstitial In species may be injected during the growth of the n-AlInP layer. Note that although the $N_D$ in the AlInP cathode layer in F is lower than in the GaAs cathode in B, the diffusion is stronger in the former (Fig. 2). This appears to be connected to the higher density of states in the conduction band of AlInP, as compared to GaAs [16], which produces a sharper gradient in the distance of Fermi level to conduction band between the surface and the bulk, and therefore a more intense imbalance and injection in group-III interstitials. The different nature of these point defects injected (Ga vs Al) could also play a role. Finally, as expected from previous observations, the diffusion mechanism partially breaks the ordered arrangement of group-III species resulting in regions with different ordering and multiple peaks in the CL emission (Fig.4, bottom).

*C. Influence of the BSF material on the diffusion process*

In the following, we assess the diffusion dynamics in Zn-doped BSF (Al)(Ga)InP alloys with different Al content susceptible to be used as BSF layers in a GaInP solar cell. In particular, BSFs of GaInP, AlGaInP and AlInP are evaluated in samples J, D and K, respectively. It must be considered that, although the amount of Zn injected during the growth of the BSF was kept constant in all the cases, the [Zn] level attained in these layers before diffusion differs from one sample to the other due to differences on the incorporation of Zn. Calibration samples grown without doped layers on top were used to measure the Zn level attained in the BSF in the absence of diffusion. The levels of [Zn] and [h+] detected by SIMS and ECV were: $2\cdot10^{18}$ cm$^{-3}$, $1\cdot10^{18}$ cm$^{-3}$ and $3\cdot10^{17}$ cm$^{-3}$ in GaInP, AlGaInP and AlInP layers. This points out important differences on the incorporation rate of Zn as the Al content in the alloy is raised. Anyway, [Zn] level in the GaInP absorber layer should be around $10^{16}$ cm$^{-3}$ in all cases, and any variation around this value can be used to detect diffusion from the BSF layer. The contrast between the profiles of [h$^+$] before and after the annealing process, shown in Fig. 5, top, reveals a strong diffusion in all samples. Note that in this case, the assessment is somehow affected by a significantly different profile between as-grown samples. In the as-grown samples, the point defects have already been injected in the structure and, therefore, the diffusion has already started during the cooling down. Consequently, the Zn element can occupy both interstitial (non-electrically active) and substitutional (active) position in the crystal lattice. This implies that the total quantity of free holes does not have to be necessarily constant during the different stages of the annealing process. This becomes particularly clear in case K, which exhibits a large difference between the doping obtained in calibration samples without diffusion, and the doping measured in the as-grown structure shown in Figure 5. Besides, the lower level of [h$^+$] detected in the annealed BSF layers as the content of Al increases, is consistent with a decrease in the solubility of Zn in AlGaInP alloys as the fraction of Al increases [17], in a similar way as the incorporation rate measured from calibration samples. This fact influences on the equilibrium [Zn] profile in the layers of the structure. In addition to a different doping level in the BSF layer, a discontinuity in [h$^+$] is observed at the BSF/GaInP interface for those BSF layers containing Al. Finally, as expected, the bandgap of the GaInP absorber layers reflects that the group-III ordering in GaInP has been almost completely disrupted in all cases (Fig.5, bottom).

### D. Incorporating barrier layers to reduce diffusion

We have shown that the Zn-redistribution takes place independently of the materials used in either the cathode or the BSF, complicating the integration of the GaInP subcell in MJSC structures. A reduction of the donor concentration in the cathode might mitigate the diffusion (as shown in Fig. 2) but would take a toll on the electric performance of the TJ, as the tunnel effect is intimately related to the doping level of the anode and the cathode of the TJ. On the other hand, the structure of a TJ typically includes two cladding layers, also termed barrier layers, designed to minimize the out-diffusion of electrically active dopant elements from the TJ to the rest of the structure [18] [19]. The MJSC structure design can take advantage of these layers to additionally hinder the out-diffusion of the injected interstitial point defects and thus reduce the Zn diffusion produced in surrounding layers.

This way, we study the effectiveness of a set of barriers, placed between the cathode and the BSF layer, with the purpose of trapping the point defects injected from the TJ cathode, (see Fig.1). The barrier layers tested consist of a 50 nm GaAs layer (D), a 500 nm GaAs layer (G), a 50 nm AlAs layer (H) and a 50 nm AlInP layer (I). The rest of the structure is kept unchanged in all samples (see Table I). The ECV profiles, presented in Fig. 6, show that a considerable reduction of the Zn diffusion can be achieved by increasing the thickness of the GaAs barrier from 50 to 500 nm (case D vs case G). The Al-based barriers like AlAs or AlInP prove to further reduce the diffusion, even for one order of magnitude less thickness. This fact may be explained by differences in the solubility and/or diffusivity of the interstitial Ga defects in the material of the barrier, or as the Al fraction varies [20]. Accordingly, the reduction of Zn diffusion implies a higher overall ordering degree in the GaInP layer: the ordered GaInP dominates the CL emission in G and, more importantly, in H and I (Fig. 7). These results demonstrate that thick n-Al(Ga)As layers could be effective to minimize the diffusion of Zn thanks to a decrease of the injection of point defects from the TJ cathode, thus enabling the integration of a highly doped TJ in a MJSC device.

### E. Impact of Zn diffusion on the perfomance of the solar cell

The diffusion of Zn may bring about substantial changes in the optoelectronic performance of a solar cell. First, the increase of the bandgap in GaInP reduces the cut-off wavelength in the QE and lowers the photogenerated current of the subcell [12]. On the other hand, the increase of the bandgap can be used to increase the open-circuit-voltage ($V_{oc}$), if the material quality is maintained. Since the ideal bandgap for the top subcell in MJSC's with 2, 3 and 4 junctions is higher than the obtained in ordered-GaInP [21], the disordering caused by a controlled Zn diffusion might be potentially beneficial. In contrast, the deviations in the electrical doping from the nominal values can take a toll on the electrical performance of the solar cell as well. For instance, the redistribution of Zn has already demonstrated to produce multiple detrimental effects on the solar cell such as the introduction electrical barriers located at the BSF/absorber interfaces [5] or the deterioration of the back-surface-passivation [22].

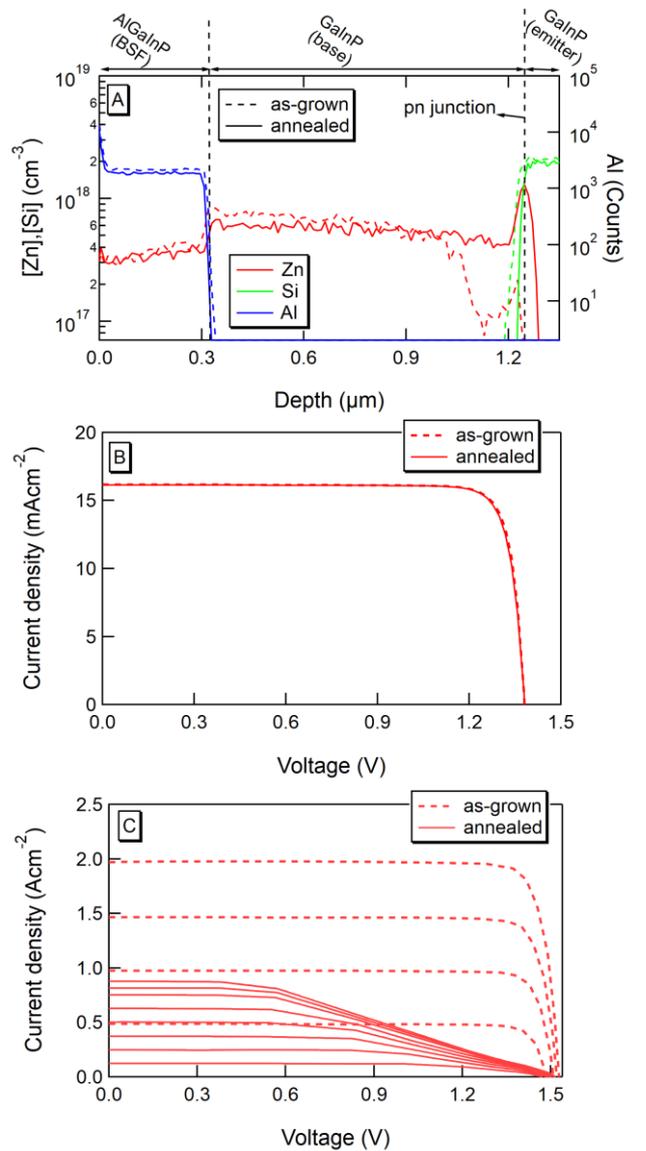

Fig. 8. For as-grown and annealed solar cells: SIMS profile of Zn, Si and Al (A); I-V curves at an irradiance around 1 sun (B); and I-V curves under concentration (C).

To evaluate the impact of Zn diffusion on a real solar cell device we compare two identical inverted GaInP solar cells (as-grown and annealed) that incorporate a complete TJ. The only difference between both samples is that the Zn profile in the annealed solar cell is varied through an ex-situ MOVPE annealing of 60 minutes at 675ºC. The growth conditions reproduce those used in the case A, aiming to promote a strong Zn diffusion across the structure. The SIMS profiles of [Zn], [Si] and [Al] are presented in Fig. 8, A.

A strong Zn diffusion can be appreciated in both cases, which is consistent with the results presented in the preceding sections. Nonetheless, a relevant difference is observed: the diffusion front in the as-grown solar cell is still comprised within the base layer, whereas in the annealed solar cell the diffusing Zn has already extended all over the base up to the *pn* junction. Hence, the incoming Zn is expected to partially counterbalance the effective n-type doping of the emitter. Consistently, the emitter sheet resistance ($R_{she}$) measured in the annealed case (1800 Ω/sq) is significantly higher than the obtained in the as-grown sample (450 Ω/sq). Since the lateral spreading of the current flow from the absorber layers towards the front metal contact occurs mainly in the emitter layer, the overall series resistance of the solar cell is intimately related to the $R_{she}$. Although the I-V curves at ~1 sun are almost identical (Fig.8, B), the impact of the series resistance becomes clear at higher irradiances (Fig.8, C). In the annealed case, the fill factor (FF) decays at very low concentrations, whereas the FF does not decay until higher $J_{sc}$ than 2.5 Acm$^{-2}$ in the as-grown sample. At this current level, the front-contact metallization may be dominating the series resistance of the device.

Therefore, in these solar cells, no noticeable internal resistive barrier is produced at the BSF/base interface, but Zn diffusion on the front layers of the structure can spoil the conductivity of the device. Although the conductivity in the emitter could be increased by modifying the thickness and/or the n-type doping of this layer, it would be desirable to prevent Zn from reaching the emitter in order to avoid additional modifications in the solar cell design. The barrier layers studied in previous sections would help attaining this.

## IV. Conclusions

The Zn diffusion dynamics in different AlGaInP/GaInP material systems corresponding to a GaInP subcell integrated into an inverted MJSC has been extensively evaluated from the perspective of the model proposed by Deppe et. al, which contemplates the assistance of point defects in the process. In particular, we have demonstrated that the growth of the cathode of the TJ after the GaInP subcell leads to an injection of point defects that induce an enhanced diffusion of Zn from the BSF to the rest of the structure. In particular, the injected group-III interstitials transfer the electrically active Zn from a substitutional to an interstitial position where the diffusivity is much higher. The injected concentration of group-III interstitial atoms depends on the electron concentration used in the cathode layer so the doping level in this layer is key to determine the intensity of the diffusion process regardless of the material used in the TJ or the Al content in the BSF. These results can be extrapolated to any (Al)(Ga)InP/GaInP system including highly doped n-type layers in the structure. We demonstrate that the diffusion of point defects can be mitigated by including Al(Ga)As or AlInP barrier layers with the proper thickness between the cathode and the BSF, which would enable the integration of highly Zn-doped BSF layers into a MJSC structure. Finally, by comparing exemplary solar cells devices with different Zn-profiles, we show that Zn diffusion can deteriorate the conductivity of the device when it reaches the emitter layer. On the other hand, no evidence of electrical barriers in the BSF/base interface were observed. These observations can help to define what could be considered a controlled Zn diffusion in the GaInP top subcell of an inverted MJSC.

### Acknowledgment

This project has been funded by the Spanish MINECO with the project TEC2017-83447-P, by the Comunidad de Madrid with the project with reference Y2018/EMT-4892 (TEFLON-CM) and by Universidad Politécnica de Madrid by Programa Propio. M. Hinojosa is funded by the Spanish MECD through a FPU grant (FPU-15/03436) and I. García is funded by the Spanish Programa Estatal de Promoción del Talento y su Empleabilidad through a Ramon y Cajal grant (RYC-2014- 15621). S. Dadgostar was funded by JCYL and FEDER (Project VA283P18).